# A bird's-eye view of scientific trading: Dependency relations among fields of science

**Erjia Yan**[1], **Ying Ding**, **Blaise Cronin**

*{eyan, dingying, bcronin}@indiana.edu*

*School of Library and Information Science, Indiana University, Bloomington, USA*

**Loet Leydesdorff**

*loet@leydesdorff.net*

*Amsterdam School of Communication Research (ASCoR), University of Amsterdam, The Netherlands*

## Abstract

We use a trading metaphor to study knowledge transfer in the sciences as well as the social sciences. The metaphor comprises four dimensions: (a) *Discipline Self-dependence*, (b) *Knowledge Exports/Imports,* (c) *Scientific Trading Dynamics*, and (d) *Scientific Trading Impact*. This framework is applied to a dataset of 221 Web of Science subject categories. We find that: (i) the Scientific Trading Impact and Dynamics of MATERIALS SCIENCE and TRANSPORTATION SCIENCE have increased; (ii) BIOMEDICAL DISCIPLINES, PHYSICS, AND MATHEMATICS are significant knowledge exporters, as is STATISTICS & PROBABILITY; (iii) in the social sciences, ECONOMICS, BUSINESS, PSYCHOLOGY, MANAGEMENT, and SOCIOLOGY are important knowledge exporters; (iv) Discipline Self-dependence is associated with specialized domains which have ties to professional practice (e.g., LAW, OPHTHALMOLOGY, DENTISTRY, ORAL SURGERY & MEDICINE, PSYCHOLOGY, PSYCHOANALYSIS, VETERINARY SCIENCES, and NURSING).

## Keywords

Scientific trading; knowledge export; knowledge import; disciplinary

## Introduction

In economics, trade refers to the transfer of ownership of goods or services from one entity to another. The notion of trade has also been applied to the transfer of things other than goods or services, such as knowledge. Since knowledge transfer cannot be observed directly (Jaffe, Trajtenberg, & Fogarty, 2000), one relies on proxy measures, notably citations. Citation linkages between articles imply a flow of knowledge from the cited to the citing entity (Jaffe, Trajtenberg, & Henderson, 1993; Van Leeuwen & Tijssen, 2000; Nomaler & Verspagen, 2008; Mehta, Rysman, & Simcoe, 2010). Several studies have used the trading metaphor to explore knowledge transfer between different disciplines (Cronin & Davenport, 1989; Cronin & Pearson, 1991; Stigler, 1994; Lockett & McWilliams, 2005; Goldstone & Leydesdorff, 2006; Cronin & Meho, 2008; Larivière, Sugimoto & Cronin, 2012; Hessey & Willett, in press).

---

[1]*Correspondence to*: Erjia Yan, School of Library and Information Science, Indiana University, 1320 E. 10th St., LI011, Bloomington, Indiana, 47405, USA. Email: eyan@indiana.edu. Tel: +1 812-606-8091. Fax: +1 812-855-6166

The application of the trading metaphor to citations takes us beyond typical bibliometric studies where the focus is on clustering disciplines (e.g., Glänzel & Schubert, 2003; Leydesdorff & Rafols, 2009; Rafols & Leydesdorff, 2009; Janssens et al., 2009; Zhang et al., 2010), mapping disciplines (e.g., Moya-Anagón et al., 2004; Leydesdorff & Rafols, 2009; Rafols, Porter, & Leydesdorff, 2010; Klavans & Boyack, 2011) or evaluating disciplines/specialties (e.g., Rinia et al., 2002; Morillo, Bordons, & Gomez, 2003; Porter et al., 2006; Bensman, 2008; Porter, Roessner, & Heberger, 2008; Zhang et al., 2010; Rafols & Meyer, 2010). Knowledge flows in the past twenty years have become more inter-sectoral, inter-organizational, inter-disciplinary, and international in character (Lewison, Rippon, & Wooding, 2005; Wagner & Leydesdorff, 2005; Autant-Bernard, Mairesse, & Massard, 2007; Ponds, Van Oort, & Frenken, 2007; Buter, Noyons, & Van Raan, 2010; Gazni, Sugimoto, & Didegah, 2011).

Perhaps the first attempt to study knowledge trading was by Xhignesse and Osgood (1967) who addressed the "problem of information exchange in psychology" (p. 778) through journal citation analysis. A receiver of information was defined as a journal whose articles cited articles published by other journals, and a source as a journal whose articles were cited by other journals. Early studies of knowledge trading were limited in terms of their scope, focusing on either a single field or small number of disciplines (Xhignesse & Osgood, 1967; Cronin & Pearson, 1991; Stigler, 1994; Lockett & McWilliams, 2005; Goldstone & Leydesdorff, 2006; Cronin & Meho, 2008; Larivière, Sugimoto & Cronin, 2012). The present study is designed to provide a comprehensive overview of trends in scientific trading across disciplines.

We begin with a formal elaboration of the trading metaphor. Figure 1 lists the terminology used in international trade (first column), the key concepts in scientific trading (second column), and their citation representations (third column). In scientific trading, for example, each *discipline* can be considered as a trading entity, the thing being transferred is *knowledge*, and the exchange currency is *citations*. A discipline *exports* its domain knowledge by *sending knowledge* through *incoming citations* and *imports* other disciplines' knowledge by *receiving knowledge* through *outgoing citations*. A discipline is thus both an *exporter* and an *importer,* to a greater or lesser extent. A *knowledge deficit* occurs if a discipline imports more knowledge than it exports; and a *knowledge surplus* occurs if a discipline exports more knowledge than it imports. Of course, not all citations are equal and thus it is impossible to quantify with precision the actual volume or heft of exported/imported knowledge. For two disciplines A and B, a *positive knowledge flow* from A to B is defined as B citing A more than A cites B; a *negative knowledge flow* is defined as B citing A less than A cites B. *Discipline self-dependence* is based on self-citation rates: a discipline is said to be independent if it has a relatively high self-citation rate and dependent if it has a relatively low self-citation rate. *Scientific Trading Impact* and *Scientific Trading Dynamics* are calculated based on the number of incoming citations (measures of impact) and the dynamics of change (Garfield, 1972; Glänzel & Moed, 2002). Note that unlike international trade, where one thing is exchanged for another, in scientific trading, knowledge is shared not given away. To be consistent with the terminology of previous studies, the term scientific trading is used here.

| Trading Metaphor | | |
|---|---|---|
| International Trade | Scientific Trading | Citation Representation |
| Country | Discipline | Subject category |
| Currency | Credit | Citation |
| Goods | Knowledge | Paper |
| Export | Sending knowledge | Cited |
| Import | Receiving knowledge | Citing |
| Exporter | Knowledge exporter/distributor | Receiving incoming citations |
| Importer | Knowledge importer/collector | Sending outgoing citations |
| Trade surplus | Knowledge surplus | Incoming citations > Outgoing citations |
| Trade deficit | Knowledge deficit | Outgoing citations > Incoming citations |
| Net capital outflow (positive) | Knowledge flow (positive) | From A to B if B cites A more than A cites B |
| Net capital outflow (negative) | Knowledge flow (negative) | From A to B if B cites A less than A cites B |
| Trading dependence | Discipline dependence | Self-citation rate |
| Trading dynamics | Scientific trading dynamics | Growth in citations |
| Trading size | Scientific trading impact | Number of citations |

Figure 1. Concepts in scientific trading

The concepts in Figure 1 are used to examine all 221 subject categories indexed in the Web of Science (WoS); subject categories are referred to as WoS categories (WCs) in version 5 of the WoS launched in August 2012 (Leydesdorff, Carley, & Rafols, in press). With this framework we are able to examine disciplinary developments and interactions in a novel fashion, in the process providing a bird's-eye view of scientific trade routes.

## Methods

### Data

Thomson Reuters' (formerly ISI's) Web of Science is one of the most comprehensive citation databases[2]. It is a well-established tool within the worlds of research evaluation and science policy making. The company assigns journals to subject categories (SCs, here also referred to as fields) based on journal-to-journal citation patterns and editorial judgment (Garfield & Pudovkin, 2002). The subject categories signify classes of specialized knowledge. Although the accuracy of the subject categories has been questioned repeatedly (e.g., Boyack, Klavans, Börner, 2005; Rafols & Leydesdorff, 2009), the categories remain the most widely used and accessible journal classification scheme. The SCs provide a clear and consistent means of tracking knowledge flows among scientific disciplines (Zitt, 2005; Van Raan, 2008). We refer the reader to Rafols and Leydesdorff (2009) for procedures on data collection. Data were harvested from the CD-ROM version of the Journal Citation Reports (JCR) of the Science Citation Index (SCI) and the Social Sciences Citation Index (SSCI) for 2007, 2008, and 2009.

At the field level, since a journal can be assigned to more than one subject category, "multiple counting" is considered in that citations from a multi-assigned journal are counted towards all assigned subject categories. "Multiple counting" avoids the arbitrariness of assigning a multi-assigned journal to either one subject category. For all fields collectively, since we calculated the total number of citations for all journals (but not fields), "multiple counting" can therefore be avoided. Thus, except for Table 1 and Table 2, the rest of the tables (including supplementary tables) use "multiple counting".

Table 1 shows the total scientific trading size for all 220 science and social sciences subject categories (221 in 2009) for the three years. For example, 30,150,625 is the number of times articles published in 2009 cited articles published in all 221 subject categories.

Table 1. Scientific Trading Impact

| Year | No. of Journals | No. of Citation links | Scientific Trading Impact (total citations) |
|---|---|---|---|
| 2007 | 7,940 | 1,460,847 | 24,979,391 |
| 2008 | 8,207 | 1,580,178 | 26,809,415 |
| 2009 | 9,216 | 1,899,373 | 30,150,625 |

### Citation flow vs. knowledge flow

---

[2] Nonetheless, there are several features of WoS that one cannot overlook: it is highly skewed towards English-language publications, covers journal articles and indexes a relatively higher number of biomedical journals.

We distinguish between citation flow and knowledge flow. Knowledge flows into a field via outgoing links and a field's own knowledge is disseminated via incoming links (Borgman & Rice, 1992; Wouters, 1998; Cronin & Meho, 2008; Leydesdorff & Probst, 2009; Buter, Noyons, & Van Raan, 2011; Leydesdorff, 2011; Levitt, Thelwall, & Oppenheim, 2011). Figure 2 provides a simple illustration of the difference between citation flow and knowledge flow. For two fields A and B, if A cites B, then A distributes citations to B (citation flow) and B diffuses knowledge to A (knowledge flow).

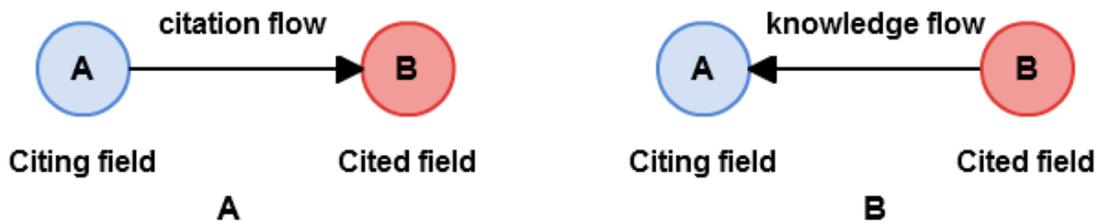

Figure 2. Citation flow vs. knowledge flow

**Scientific disciplines: dimensions and characteristics**

Disciplinary trading can be captured along four dimensions: (i) *Discipline Self-dependence*, (ii) *Knowledge Exports/Imports*, (iii) *Scientific Trading Dynamics*, and (iv) *Scientific Trading Impact*. The interrelationship of these dimensions is shown in the *Discussion* section. Figure 3 provides visualizations. For two fields A and B,

- *Discipline Self-dependence* is represented by the size of the inner circle area; the higher the self-citation rate, the greater the area (Figure 3A).
- *Knowledge Exports/Imports* are represented by the size of the outgoing and incoming arrows. Arrow size indicates the amount of imported and exported knowledge (Figure 3B).
- Circles represent *Scientific Trading Dynamics*. The circle area size mirrors the amount of knowledge exported by a discipline in different years (Figure 3C).
- *Scientific Trading Impact* shows the sum of exported knowledge for different fields during the same time period (Figure 3D).

To illustrate, we provide two examples. The first compares two fields in the same year: Field A has a lower Scientific Trading Impact, higher Discipline Self-dependence, and a lower Export/Import ratio than field B (Figure 3E). The second compares the same field over time: the field at the second point in time has a higher Scientific Trading Impact, a lower Discipline Self-dependence, and a lower Export/Import ratio than it had at the first (Figure 3F). We applied these four dimensions to the 221 subject categories to reveal the nature and scale of knowledge trading across disciplines.

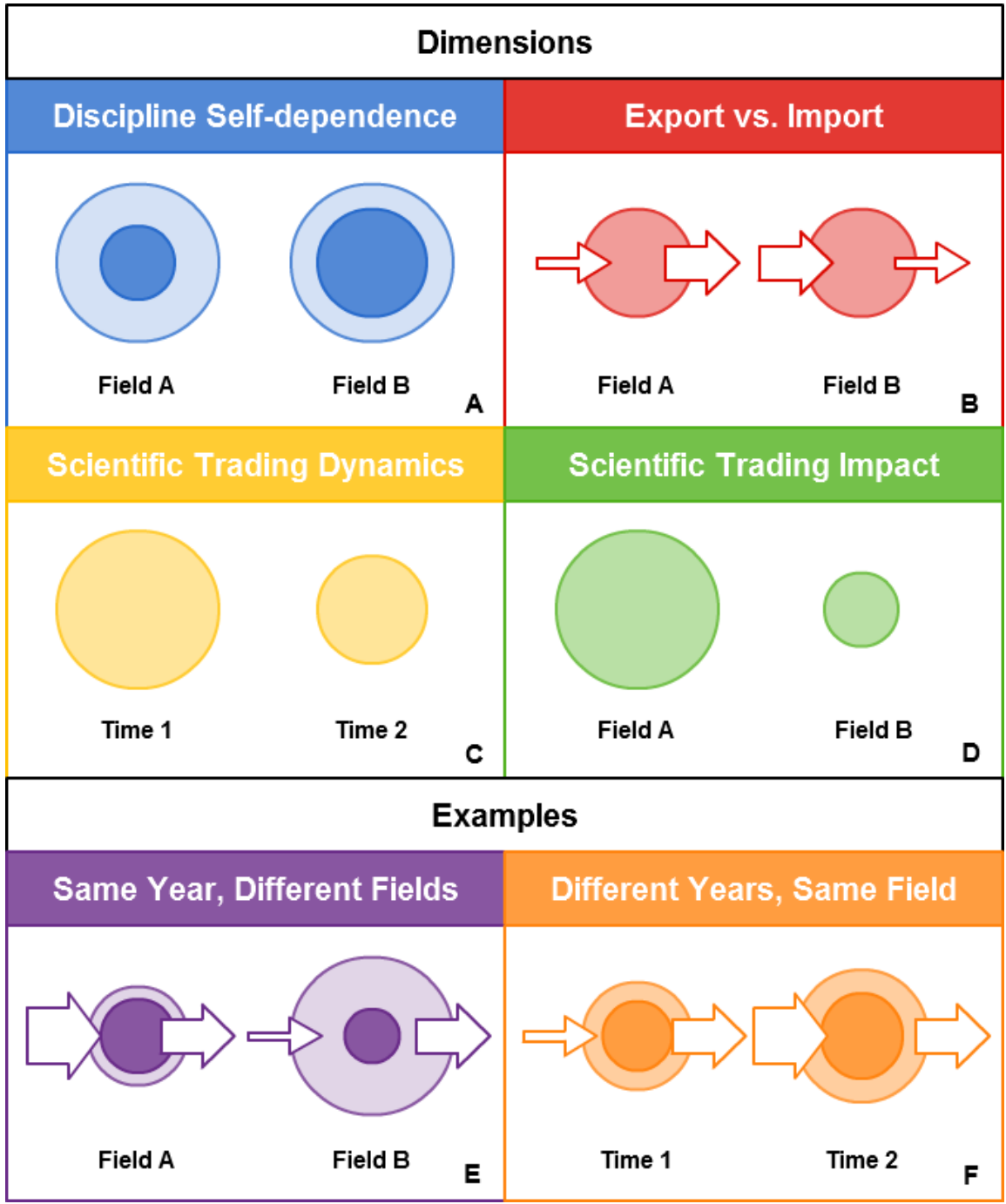

Figure 3. Disciplinary dimensions

## Results

**The acceleration of science (Scientific Trading Impact and Dynamics)**

Human knowledge, in the form of scholarly publications, is growing apace. Earlier knowledge trading studies did not take the acceleration of science into account. As more and more scientific papers are published each year, the amount of citable (exportable) knowledge increases. In other words, a field may show a year-on-year increase in exports, but that growth rate may actually be less than the rate at which the overall citable (exportable) knowledge stock is increasing.

Table 2. Scientific Trading Impact and Dynamics[3]

| Year range | Increment of citable (exportable) knowledge | No. of fields that export knowledge more than the increment | No. of fields that export knowledge less than the increment |
|---|---|---|---|
| 2007~2008 | 7.33% | 144 | 77 |
| 2008~2009 | 12.46% | 125 | 96 |
| 2007~2009 | 20.70% | 133 | 88 |
| Union of 2007~2008 and 2008~2009 | | 98 | 61 |

The overall acceleration rate of science is shown in Table 2. Total Scientific Trading Impact increased by 20.70% from 2007 to 2009. A hundred and thirty three fields exported knowledge at a rate greater than the overall increment. Ninety-eight fields exported more knowledge than the overall acceleration rate in both time periods – from 2007 to 2008 and from 2008 to 2009, while 61 exported less. The former have, in our terms, an increased Scientific Trading Impact, the latter a reduced Scientific Trading Impact.

Figure 4 illustrates the distribution of Scientific Trading Dynamics for all 221 fields. The distribution did not pass the normal significance test with Kolmogorov-Smirnov's asymptotic significance equal to 0.004 ($p<0.05$). The distribution is positively skewed (statistic is 1.2 which is more than three times larger than its standard error 0.16) and leptokurtic (statistic is 4.65 which is more than three times larger than its standard error 0.33). The mean for the distribution is 0.31 and standard deviation is 0.24.

---

[3] MULTIDISCIPLINARY SCIENCES is excluded, as it is not considered a discipline.

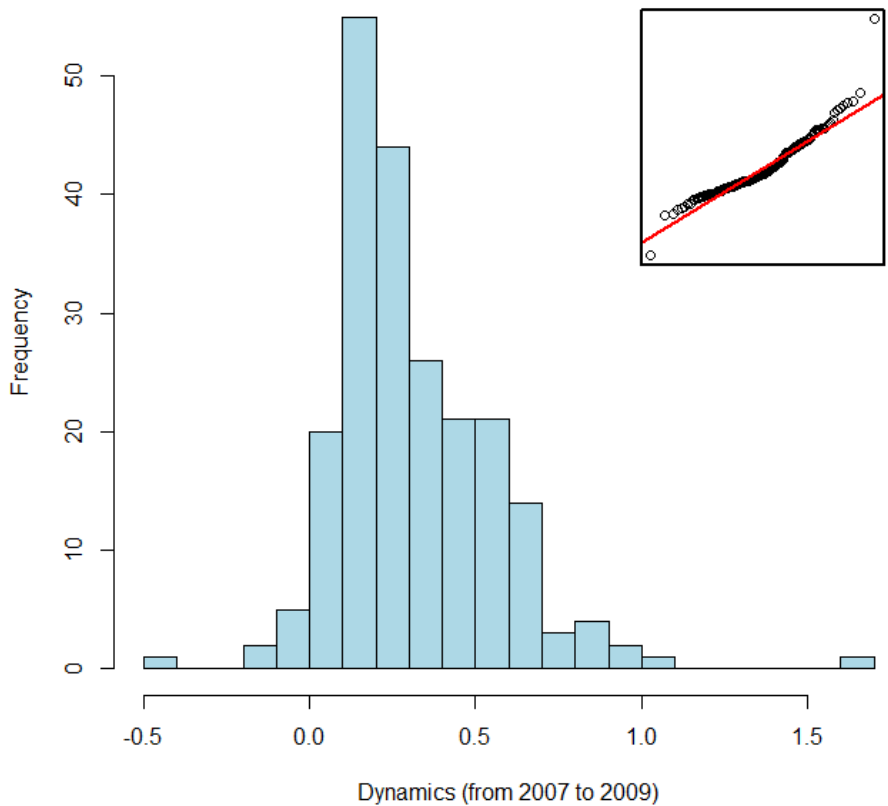

Figure 4. Distribution of Scientific Trading Dynamics (from 2007 to 2009) with Normal Quantile-Quantile Plot

Tables 3 and 4 show the top-10 fields ranked by their Scientific Trading Dynamics. "C" denotes the number of incoming citations (exported knowledge) and "P" the number of publications. Dynamic export changes for all 221 fields are included in the supplementary materials (Supplementary Table 1). The rate of increase should be used for benchmarking purposes: a research unit (be it an author, journal, or institution) in a given field may have increased its rate of knowledge exports but less than its field average, in which case it is lagging behind the overall pace of scientific development of its field.

Table 3. Top-10 fields based on export increases (2007-2009)

| Subject categories* | | Exports in 2007 | Exports in 2009 | Increase (%) |
|---|---|---|---|---|
| NANOSCIENCE & NANOTECHNOLOGY | C | 241,198 | 646,645 | 168.10% |
| | P | 10,000 | 17,747 | 77.47% |
| MATERIALS SCIENCE, BIOMATERIALS | C | 92,103 | 184,584 | 100.41% |
| | P | 2,665 | 4,159 | 56.06% |
| TRANSPORTATION SCIENCE & TECHNOLOGY | C | 19,330 | 37,110 | 91.98% |
| | P | 1,321 | 2,394 | 81.23% |
| HEALTH POLICY & SERVICES | C | 10,101 | 19,331 | 91.38% |
| | P | 2,499 | 3,330 | 33.25% |
| PSYCHOLOGY, MATHEMATICAL | C | 8,584 | 16,156 | 88.21% |
| | P | 506 | 596 | 17.79% |
| AGRICULTURAL ENGINEERING | C | 37,841 | 70,511 | 86.33% |
| | P | 1,351 | 2,103 | 55.66% |
| SOCIAL SCIENCES, BIOMEDICAL | C | 25,874 | 47,574 | 83.87% |
| | P | 1,793 | 1,989 | 10.93% |
| TRANSPORTATION | C | 7,963 | 14,493 | 82.00% |
| | P | 757 | 873 | 15.32% |
| MATERIALS SCIENCE, MULTIDISCIPLINARY | C | 1,080,221 | 1,890,081 | 74.97% |
| | P | 40,905 | 51,853 | 26.76% |
| ETHICS | C | 7,415 | 12,833 | 73.07% |
| | P | 1,049 | 1,501 | 43.09% |

*C: number of citations; P: number of publications

The Scientific Trading Impact of a field tells us whether its domain knowledge is recognized and valued – the idea of citations as endorsements (Merton, 1968; Cronin, 1984). We can see from Table 3 that materials science (including NANOSCIENCE & NANOTECHNOLOGY, MATERIALS SCIENCE, BIOMATERIALS, and MATERIALS SCIENCE, MULTIDISCIPLINARY) and transportation (including TRANSPORTATION SCIENCE & TECHNOLOGY and TRANSPORTATION) have increased their Scientific Trading Impact in recent years, which means that they are becoming more visible and valued by other fields.

Table 4. Top-10 fields in terms of reduced exports (2007-2009)

| Subject categories* | | Exports in 2007 | Exports in 2009 | Change (%) |
|---|---|---|---|---|
| PSYCHOLOGY, BIOLOGICAL | C | 2,699 | 1,369 | -49.28% |
| | P | 1,014 | 1,149 | 13.31% |
| MEDICAL ETHICS | C | 8,014 | 7,028 | -12.30% |
| | P | 416 | 592 | 42.31% |
| HISTORY & PHILOSOPHY OF SCIENCE | C | 14,033 | 12,418 | -11.51% |
| | P | 1,007 | 1,162 | 15.39% |
| PSYCHOLOGY | C | 335,943 | 310,887 | -7.46% |
| | P | 3,984 | 4,923 | 23.57% |
| PSYCHOLOGY, PSYCHOANALYSIS | C | 7,153 | 6,755 | -5.56% |
| | P | 467 | 438 | -6.21% |
| MEDICAL LABORATORY TECHNOLOGY | C | 86,805 | 82,734 | -4.69% |
| | P | 2,559 | 2,707 | 5.78% |
| STATISTICS & PROBABILITY | C | 269,364 | 262,307 | -2.62% |
| | P | 6,512 | 6,844 | 5.10% |
| ORNITHOLOGY | C | 33,466 | 32,680 | -2.35% |
| | P | 1,135 | 956 | -15.77% |
| AGRICULTURAL ECONOMICS & POLICY | C | 9,130 | 9,317 | 2.05% |
| | P | 424 | 549 | 29.48% |
| PSYCHIATRY | C | 692,904 | 790,573 | 2.19% |
| | P | 10,258 | 11,829 | 15.31% |

*C: number of citations; P: number of publications

By way of contrast, the 10 fields listed in Table 4 have acceleration rates below the overall rate, which may mean that they are becoming somewhat less central, influential, or popular among other fields of science. Note, too, that some closely related fields appear in both Table 3 (e.g., PSYCHOLOGY, MATHEMATICAL and ETHICS) and Table 4 (e.g., PSYCHOLOGY, BIOLOGICAL, PSYCHOLOGY and PSYCHOLOGY, PSYCHOANALYSIS). This would seem to suggest that sub-field analysis is needed to ensure that we don't draw false conclusions about a field's waxing and waning. For some fields (e.g., NANOSCIENCE & NANOTECHNOLOGY and MATERIALS SCIENCE, BIOMATERIALS), the increased Scientific Trading Impact is due largely to the increase in the number of publications; for other fields (e.g., PSYCHOLOGY, MATHEMATICAL and SOCIAL SCIENCES, BIOMEDICAL) the increase can be attributed to greater per publication impact (as reflected in citation counts).

**Exporters and importers**

If a field's outgoing citations exceed its incoming citations, it's a net importer of knowledge; if incoming citations exceed outgoing citations, it's a net exporter (see Table 5).

Table 5. Exporter vs. importer

| Role | 2007 | 2008 | 2009 |
|---|---|---|---|
| Export>Import | 79 | 88 | 81 |
| Import>Export | 141 | 132 | 140 |

Figure 5 shows the distribution of Export/Import Ratios for all 221 fields for 2009. The distribution follows a normal distribution with Kolmogorov-Smirnov's asymptotic significance equal to 0.18 (p>0.05). The mean for the distribution is 0.95 and the standard deviation is 0.19. The results indicate that a majority of fields have a balanced knowledge economy, and only a few play a salient role as knowledge importers (ratio below 0.77) or knowledge exporters (ratio above 1.13).

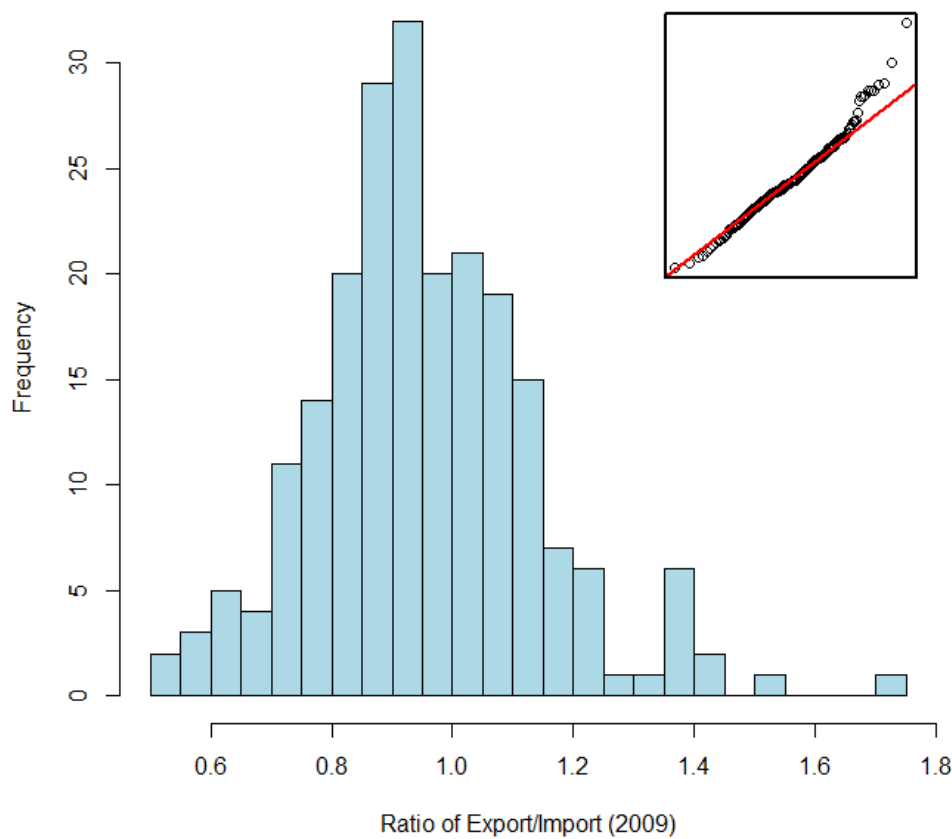

Figure 5. Distribution of Export/Import Ratios (2009) with Normal Quantile-Quantile Plot

The marketplace of ideas is largely barrier-free—we leave aside here issues relating to tolls and open access publishing. Since knowledge is traded freely, successful exporters are most likely to be those scholars with the highest quality and/or most useful goods to offer. Table 6 lists the most significant exporters in the sciences and social sciences. These fields are the engines that power science and technology development. For example, MEDICINE, GENERAL & INTERNAL, HEMATOLOGY, CELL BIOLOGY, and PERIPHERAL VASCULAR DISEASE power biomedical research; STATISTICS & PROBABILITY and COMPUTER SCIENCE, HARDWARE & ARCHITECTURE, and PHYSICS, MULTIDISCIPLINARY power physical, mathematical, and engineering research; SOCIAL SCIENCES, MATHEMATICAL METHODS, and PSYCHOLOGY, SOCIAL power social sciences research. Export/import ratios for all 221 fields are included in the supplementary material (Supplementary Table 2).

Table 6. Top-10 significant exporters based on export/import ratios (2009)

| Subject categories | Imports | Exports | Exports/imports |
| --- | --- | --- | --- |

| Subject category | Imports | Exports | Exports/imports |
|---|---|---|---|
| MEDICINE, GENERAL & INTERNAL | 635,835 | 1,100,441 | 1.73 |
| STATISTICS & PROBABILITY | 171,231 | 262,307 | 1.53 |
| SOCIAL SCIENCES, MATHEMATICAL METHODS | 14,974 | 21,421 | 1.43 |
| COMPUTER SCIENCE, HARDWARE & ARCHITECTURE | 58,952 | 83,897 | 1.42 |
| PSYCHOLOGY, MATHEMATICAL | 11,596 | 16,156 | 1.39 |
| HEMATOLOGY | 564,113 | 785,838 | 1.39 |
| PSYCHOLOGY, SOCIAL | 111,371 | 155,045 | 1.39 |
| PHYSICS, MULTIDISCIPLINARY | 714,276 | 978,353 | 1.37 |
| CELL BIOLOGY | 1,407,519 | 1,918,916 | 1.36 |
| PERIPHERAL VASCULAR DISEASE | 478,743 | 651,799 | 1.36 |

The fields in Table 7 are significant importers. Their export/import ratios are less than one, indicating a knowledge trading deficit.

Table 7. Top-10 significant importers based on export/import ratios (2009)

| Subject categories | Imports | Exports | Exports/imports |
|---|---|---|---|
| ETHICS | 25,144 | 12,833 | 0.51 |
| BIOLOGY | 666,711 | 354,614 | 0.53 |
| HEALTH POLICY & SERVICES | 34,627 | 19,331 | 0.56 |
| NURSING | 102,797 | 58,721 | 0.57 |
| PARASITOLOGY | 186,396 | 110,826 | 0.59 |
| MATERIALS SCIENCE, CHARACTERIZATION & TESTING | 34,469 | 20,960 | 0.61 |
| MEDICAL ETHICS | 11,295 | 7,028 | 0.62 |
| INFORMATION SCIENCE & LIBRARY SCIENCE | 19,987 | 12,728 | 0.64 |
| INTEGRATIVE & COMPLEMENTARY MEDICINE | 51,941 | 33,208 | 0.64 |
| VETERINARY SCIENCES | 395,908 | 256,961 | 0.65 |

*Knowledge flow and knowledge surplus*

We next zoom in to capture the extent to which any given subject category is a net exporter to other SCs; that is to say, we identify the fields that have the largest number of trading partners. Table 8 lists the top-10 science SCs and the top-10 social sciences SCs. Table 9 lists the top-10 science subject categories and top-10 social sciences subject categories based on (positive) knowledge surplus.

Table 8. Top-10 science and social sciences exporters

| Science subject categories | No. of SCs | Social sciences subject categories | No. of SCs |
|---|---|---|---|
| STATISTICS & PROBABILITY | 211 | ECONOMICS | 153 |
| MEDICINE, GENERAL & INTERNAL | 196 | PSYCHOLOGY, SOCIAL | 139 |
| BIOCHEMISTRY & MOLECULAR BIOLOGY | 186 | BUSINESS | 138 |
| MATHEMATICAL & COMPUTATIONAL BIOLOGY | 177 | SOCIOLOGY | 136 |

| GENETICS & HEREDITY | 170 | SOCIAL SCIENCES, MATHEMATICAL METHODS | 134 |
| CELL BIOLOGY | 169 | MANAGEMENT | 120 |
| PERIPHERAL VASCULAR DISEASE | 166 | PSYCHOLOGY, CLINICAL | 117 |
| CARDIAC & CARDIOVASCULAR SYSTEMS | 166 | PSYCHOLOGY, EDUCATIONAL | 115 |
| HEMATOLOGY | 162 | PSYCHOLOGY, EXPERIMENTAL | 115 |
| PSYCHOLOGY | 161 | PSYCHOLOGY, MULTIDISCIPLINARY | 106 |

Table 9. Top-10 science and social sciences SCs in terms of knowledge surplus

| Science subject categories | Knowledge surplus | Social sciences subject categories | Knowledge surplus |
| --- | --- | --- | --- |
| BIOCHEMISTRY & MOLECULAR BIOLOGY | 898,501 | ECONOMICS | 75,406 |
| CELL BIOLOGY | 526,045 | BUSINESS | 48,305 |
| MEDICINE, GENERAL & INTERNAL | 465,714 | PSYCHOLOGY, SOCIAL | 47,841 |
| CHEMISTRY, MULTIDISCIPLINARY | 431,493 | MANAGEMENT | 37,653 |
| PHYSICS, MULTIDISCIPLINARY | 279,577 | SOCIOLOGY | 27,741 |
| GENETICS & HEREDITY | 247,151 | PSYCHOLOGY, CLINICAL | 19,827 |
| ONCOLOGY | 238,465 | BUSINESS, FINANCE | 16,465 |
| HEMATOLOGY | 235,656 | PSYCHOLOGY, EDUCATIONAL | 15,168 |
| IMMUNOLOGY | 230,924 | PSYCHOLOGY, DEVELOPMENTAL | 12,901 |
| PHYSICS, CONDENSED MATTER | 210,996 | PSYCHOLOGY, EXPERIMENTAL | 12,508 |

STATISTICS & PROBABILITY functions as a knowledge exporter to almost all other subject categories (211 in total). It provides useful methodological applications to many scientific and social science disciplines. As for knowledge surplus, the top-10 SCs in science comprise the biomedical sciences, chemistry, and physics. In the social sciences, ECONOMICS is the leading knowledge exporter, followed by BUSINESS, PSYCHOLOGY, MANAGEMENT, and SOCIOLOGY. There are at least two possible interpretations. First, these fields may be more mature than other social sciences disciplines and thus attract the attention of less established, less cognitively assured fields. Second, these SCs tend to preferentially cite their own literature because they have less permeable boundaries and less need to borrow tools, methods or theories from other fields (see (Klein, 1996) on disciplinary boundaries and boundary crossing). The top social science fields measured by knowledge surplus include economics, business, management, and psychology.

*Science hubs*

Hubs are the most connected nodes in a network. In terms of scientific trading, the fields with the largest volume of exported and imported knowledge are hubs. Similar to the busiest airports in the world, science hubs interconnect smaller fields. Two fields may not have direct trading connections, but they can always link indirectly via hubs. These science hubs function as knowledge processors, absorbing and distributing value-added knowledge to other fields.

Figure 6 and Figure 7 show the distributions of the sizes of imported and exported knowledge for all 221 fields for 2009. The two distributions display a noticeable power-law distribution pattern (Kolmogorov-Smirnov's asymptotic significance equals zero) in that most fields are limited in the size of imported or exported knowledge and only a few fields function as science hubs.

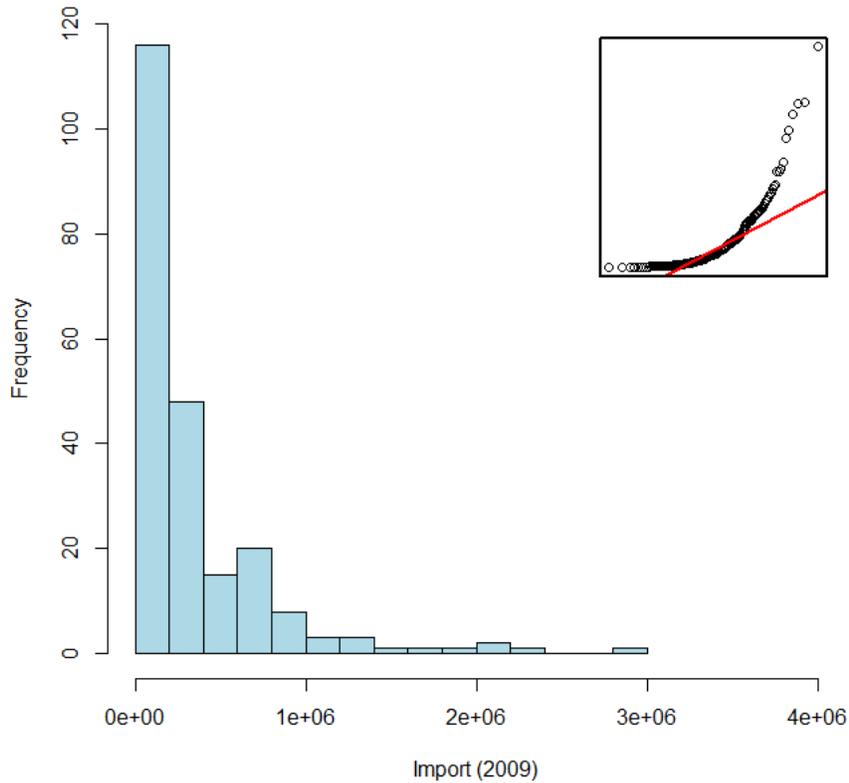

Figure 6. Distribution of Imports (2009) with Normal Quantile-Quantile Plot

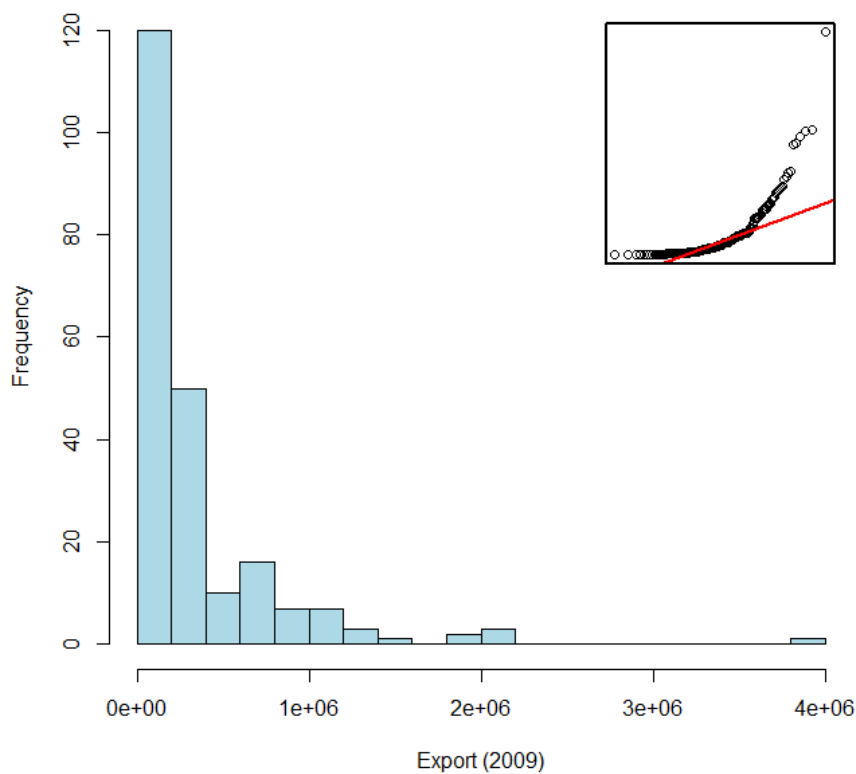

Figure 7. Distribution of Exports (2009) with Normal Quantile-Quantile Plot

The top-10 science hubs are shown in Table 10.

Table 10. Top-10 science hubs

| Name | Imports | Exports | Imports+exports |
|---|---|---|---|
| BIOCHEMISTRY & MOLECULAR BIOLOGY | 2,969,848 | 3,820,866 | 6,790,714 |
| CHEMISTRY, PHYSICAL | 2,196,028 | 2,029,046 | 4,225,074 |
| NEUROSCIENCES | 2,062,891 | 2,129,998 | 4,192,889 |
| MATERIALS SCIENCE, MULTIDISCIPLINARY | 2,216,953 | 1,890,081 | 4,107,034 |
| CHEMISTRY, MULTIDISCIPLINARY | 1,843,270 | 2,140,721 | 3,983,991 |
| CELL BIOLOGY | 1,407,519 | 1,918,916 | 3,326,435 |
| PHARMACOLOGY & PHARMACY | 1,745,816 | 1,339,498 | 3,085,314 |
| PHYSICS, APPLIED | 1,294,934 | 1,428,746 | 2,723,680 |
| ONCOLOGY | 1,294,057 | 1,397,783 | 2,691,840 |
| ENVIRONMENTAL SCIENCES | 1,327,559 | 1,150,162 | 2,477,721 |

According to Lenoir (1997), the creation of domain knowledge reflects the dominant relations of economic, social, and political power in society. The prominence of biomedical fields as science hubs underscores his point: biomedical products have high market value (economic power), are important to the well-being of society (social power), and are sensitive to science policy decisions (political power). The importance of this SC is reflected in the magnitude of the funds allocated to biomedical research every year and the vastness of the research literature.

**Discipline Self-dependence**

Disciplines vary greatly in terms of their permeability, cognitive autonomy and self-dependence (Klein, 1996). Some are interdisciplinary and porous while others are self-contained and have a quite distinct core. Bourdieu (1977), for instance, considered sociology to be less autonomous than biomedical fields, and the latter less autonomous than physics. The assumption is that an independent field preferentially cites its own publications; the higher the rate of self-citation, the more self-contained a field is (e.g., Borgman & Rice, 1992; Rinia et al., 2002; Guerrero et al., 2007; Cronin & Meho, 2008; Porter & Rafols, 2009; Leydesdorff & Probst, 2009; Rafols & Meyer, 2010; Lillquist & Green, 2010; Buter, Noyons, & Van Raan, 2011; Leydesdorff, 2011; Levitt, Thelwall, & Oppenheim, 2011).

Table 11. Discipline Self-dependence

|  | 2007 | 2008 | 2009 |
|---|---|---|---|
| Primarily depends on itself | 174 | 171 | 161 |
| Primarily depends on others | 46 | 49 | 60 |

It can be seen from Table 11 that most fields are predominantly self-citers, i.e., self-reliant, even though self-dependence is decreasing overall. Some, however, depend on sources other than themselves for ideas. Several factors come into play. It is possible that a new or emerging field draws significantly on the literatures of more established fields as it develops and consolidates its own intellectual base (Larivière, Sugimoto, & Cronin, 2012). A smaller field may also lie at the intersection of two (or several) larger fields and such an interstitial field may draw heavily on the bookending fields (Leydesdorff & Probst, 2009).

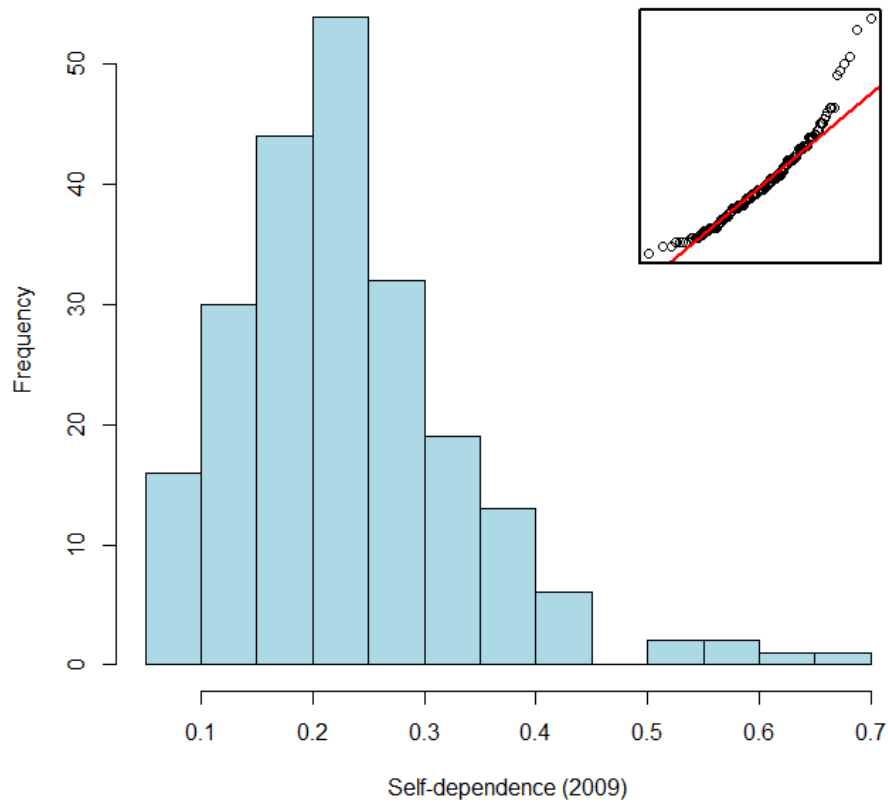

Figure 8. Distribution of Self-dependence (2009) with Normal Quantile-Quantile Plot

Figure 8 shows the distribution of Self-dependence for all 221 fields for 2009. The distribution did not pass the normal significance test with Kolmogorov-Smirnov's asymptotic significance equal to 0.002 ($p<0.05$). The distribution is positively skewed (statistic is 1.2 which is more than three times larger than its standard error 0.16) and leptokurtic (statistic is 2.68 which is more than three times larger than its standard error 0.33). The mean for the distribution is 0.23 and the standard deviation is 0.10, suggesting that although most fields are predominantly self-citers, the combination of cited knowledge from all other fields may account for a higher percentage.

Table 12 shows the most independent disciplines in both the sciences and social sciences. Self-citation ratios for all 221 fields are included in the supplement material (Supplementary Table 2).

Table 12. Top-10 independent disciplines (2009)

| Name | Exports | Self-citations | Ratio |
|---|---|---|---|
| LAW | 61,809 | 42,025 | 0.68 |
| ASTRONOMY & ASTROPHYSICS | 686,519 | 448,998 | 0.65 |
| OPHTHALMOLOGY | 232,581 | 134,928 | 0.58 |
| DENTISTRY, ORAL SURGERY & MEDICINE | 206,287 | 115,870 | 0.56 |

| MATHEMATICS | 307,136 | 164,870 | 0.54 |
| PSYCHOLOGY, PSYCHOANALYSIS | 6,755 | 3,599 | 0.53 |
| VETERINARY SCIENCES | 256,961 | 114,065 | 0.44 |
| LINGUISTICS | 37,672 | 16,654 | 0.44 |
| NURSING | 58,721 | 25,714 | 0.44 |
| EDUCATION & EDUCATIONAL RESEARCH | 65,164 | 27,790 | 0.43 |

Each is highly specialized and strongly reliant on its own knowledge base. Specialized disciplines include vocational and professional fields, for example, LAW, OPHTHALMOLOGY, DENTISTRY, ORAL SURGERY & MEDICINE, PSYCHOLOGY, PSYCHOANALYSIS, VETERINARY SCIENCES, and NURSING. These disciplines typically have institutional arrangements (formal educational programs, faculties, accrediting agencies, learned/professional societies, etc.) and paraphernalia (meetings, textbooks, scholarly journals, prizes, etc.) that newer, small, less established fields may lack or be unable to put in place.

Table 13. Top-10 dependent disciplines (2009)

| Name | Exports | Self-citations | Ratio |
|---|---|---|---|
| PSYCHOLOGY, BIOLOGICAL | 1,369 | 66 | 0.05 |
| SOCIAL SCIENCES, MATHEMATICAL METHODS | 21,421 | 1,419 | 0.07 |
| MEDICINE, RESEARCH & EXPERIMENTAL | 615,899 | 43,815 | 0.07 |
| BIOLOGY | 354,614 | 28,967 | 0.08 |
| ANATOMY & MORPHOLOGY | 50,740 | 4,195 | 0.08 |
| MICROSCOPY | 29,219 | 2,436 | 0.08 |
| PSYCHOLOGY, MATHEMATICAL | 16,156 | 1,348 | 0.08 |
| BIOPHYSICS | 672,529 | 56,720 | 0.08 |
| LIMNOLOGY | 107,229 | 9,327 | 0.09 |
| ANDROLOGY | 11,723 | 1,041 | 0.09 |

Most dependent disciplines (see Table 13) are located at the intersection of two or more main disciplines. For example, PSYCHOLOGY, BIOLOGICAL is at the intersection of PSYCHOLOGY and BEHAVIORAL SCIENCES, and thus it cites these two fields most (more than 40%) but does not cite many of its own publications (less than 5%). BIOLOGY is a well-established field but since the second half of the twentieth century it has become increasingly dependent on BIOCHEMISTRY & MOLECULAR BIOLOGY (23%), followed by CELL BIOLOGY (14%), NEUROSCIENCES (11%), GENETICS & HEREDITY (10%), and BIOLOGY (8%).

## Discussion

### Interrelationship of the scientific trading dimensions

In order to test the reliability of the scientific trading dimensions, Spearman's rank correlation coefficients were calculated for Scientific Trading Dynamics, Ratio of Export/Import, Import, Export, and Self-dependence for 2009, with scatter plot visualizations (Figure 9).

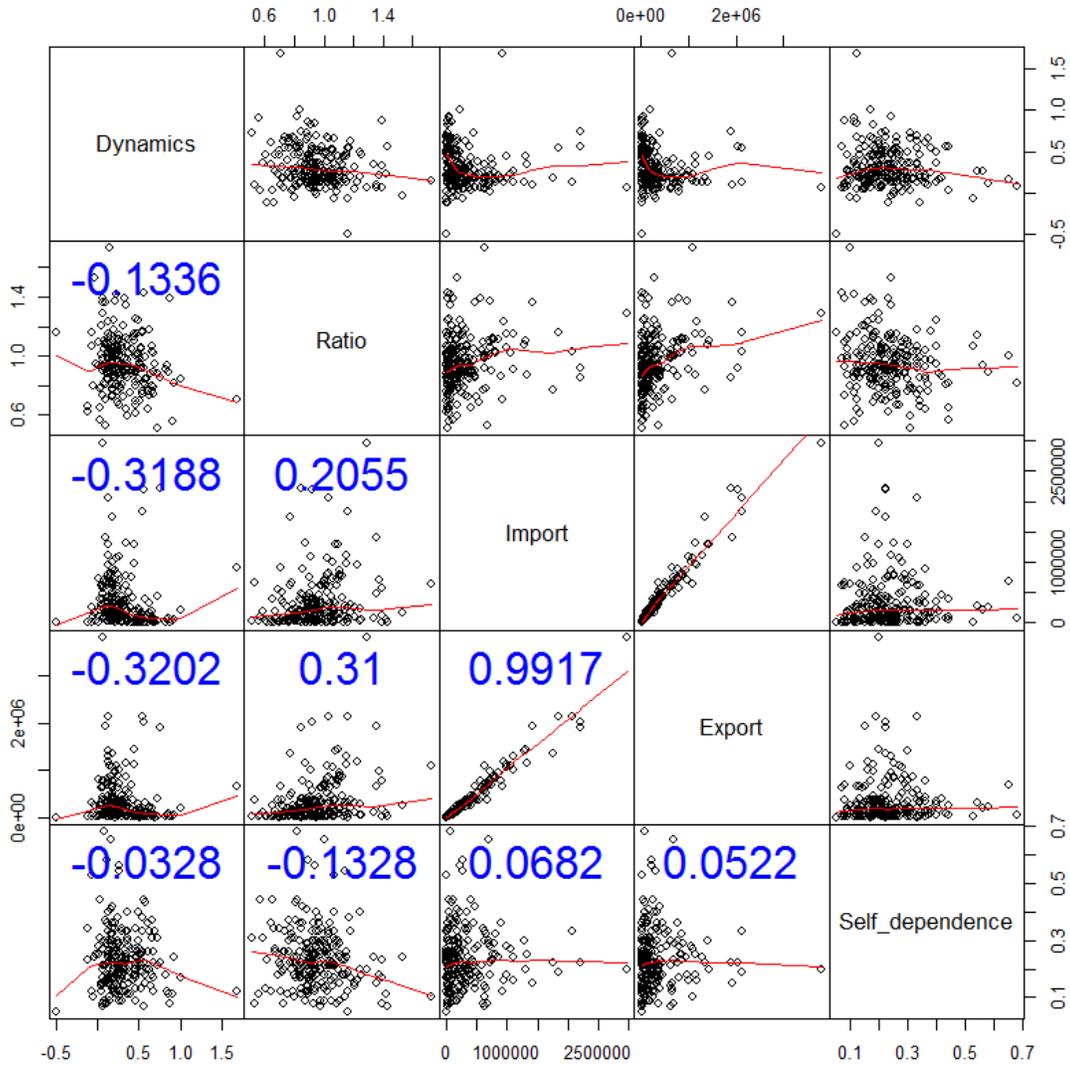

Figure 9. Spearman's rank correlation coefficients and scatter plots for pairs of indicators

In Figure 9, except for the correlation coefficient ($R^2$=0.9917) between Import and Export, the strength of the other correlation coefficients ranges from mild to weak. This indicates that the dimensions are not susceptible to multicollinearity and that each one works independently and is thus capable of describing a particular characteristic of disciplines. Combined, the dimensions provide a comprehensive perspective on disciplinary characteristics.

**Applying the scientific trading dimensions to a variety of fields**

Using the dimensions we introduced, we highlight ten characteristic types and illustrate each one with a number of SCs (Figure 10). In general, scientific fields have a higher *Scientific Trading Impact* than social science fields. Biomedical sciences (BIOCHEMISTRY & MOLECULAR BIOLOGY and MEDICINE, GENERAL & INTERNAL), physics, and chemistry have higher levels of dependence and higher *Export/Import* ratios. Interdisciplinary science fields, such as CHEMISTRY, PHYSICS, MATERIAL SCIENCE, MULTIDISCIPLINARY, and ENVIRONMENTAL SCIENCES, have lower *Discipline Self-dependence* and lower *Export/Import* ratios. Specialized biomedical sciences, such as NEUROSCIENCES and ONCOLOGY have higher *Discipline Self-dependence* and higher *Export/Import* ratios. The social sciences (e.g., INFORMATION SCIENCE & LIBRARY SCIENCE, LAW, EDUCATION & EDUCATIONAL RESEARCH, POLITICAL SCIENCE, MANAGEMENT, and BUSINESS) tend to have higher levels of *Discipline Self-dependence.*

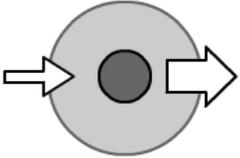

Figure 10. Characteristics of subject categories

## Characteristics of the sciences and the social sciences

We visualized the distributions for trading dynamics (Figure 4), export/import ratios (Figure 5), import (Figure 6), export (Figure 7), and self-dependence (Figure 8); and found that except for the distribution of export/import ratios, the rest of the trading dimensions did not follow a normal distribution — in particular, the import and export distributions displayed a power law pattern. Such power law properties have been found for different research entities, such as paper citations (e.g., Redner, 1998), author productivity (e.g., Lotka, 1926), journal productivity (e.g., Bradford, 1934), or institutional citations (e.g., Carvalho & Batty, 2006; Yan & Sugimoto, 2011), i.e., only a limited number of these entities have high scientific productivity and/or impact while the majority have low scientific productivity and/or impact. In our study, at the field level, there is an apparent difference between the science and the social science in their scientific trading impact (see Tables 9 and 10). Given this difference, do the social sciences, due to their size, suffer from the preferential attachment effect (Merton, 1968; Barabási & Albert, 1999)? To address this question, we calculated the increment rate for all social science fields as well as the percentage of trading impact for all social science fields in Table 14.

Table 14. Scientific Trading Impact and Dynamics for the social sciences

| Year Range | Increment of citable knowledge for all fields (from Table 2) | Increment of citable knowledge for social science fields | Year | Percentage of trading impact for all social science fields against all fields |
|---|---|---|---|---|
| 2007~2008 | 7.33% | 15.4% | 2007 | 3.15% |
| 2008~2009 | 12.46% | 30.51% | 2008 | 3.34% |
| 2007~2009 | 20.70% | 50.61% | 2009 | 3.84% |

Table 14 shows that the knowledge increment rates for social science fields grew faster than for all fields, while the percentage of trading impact for all social science fields have increased from 3.15% in 2007 to 3.84% in 2009. The results suggest that despite the power law pattern for knowledge import and export distributions, social science fields may not be susceptible to the preferential attachment effect in that these fields are becoming more visible by exporting more knowledge and having a higher share of scientific trading. The results indicate that the sciences and social sciences are quite different (Suppes, 1984; Nederhof, 2006). In future research we intend to develop fine-grained indicators to study in more detail how knowledge is disseminated.

## Conclusion

We developed a set of concepts to describe scientific trading. Using these concepts, we fashioned a framework comprising four dimensions: (a) *Discipline Self-dependence*, (b) *Exports/Imports*, (c) *Scientific Trading Dynamics*, and (d) *Scientific Trading Impact*. This framework enabled us to develop a unique, data-rich bird's-eye view of trends in knowledge trading between disciplines and fields. Our study reveals the permeability and self-sufficiency of different scientific and social science disciplines. The findings should stimulate further research into the nature and dynamics of disciplinarity and interdisciplinarity and also help inform science policy making.

## Acknowledgement

We are grateful to Thomson Reuters for providing access to the data and the reviewers for their constructive feedback.

## References

Autant-Bernard, C., Mairesse, J., & Massard, N. (2007). Spatial knowledge diffusion through collaborative networks. *Papers in Regional Science, 86*(3), 341-350.

Barabási, A.-L., & Albert, R. (1999). Emergence of scaling in random networks. *Science, 286*(5439), 509-512.

Bensman, S. J. (2008). Distributional differences of the impact factor in the sciences versus the social sciences: An analysis of the probabilistic structure of the 2005 journal citation reports. *Journal of the American Society for Information Science and Technology, 59*(9), 1366-1382.

Borgman, C. L., & Rice, R. E. (1992). The convergence of information science and communication: A bibliometric analysis. *Journal of the American Society for Information Science, 43*(6), 397-411.

Bourdieu, P. (1977). *Outline of a theory of practice*. Cambridge, United Kingdom: Cambridge University Press.

Boyack, K. W., Klavans, R., & Börner, K. (2005). Mapping the backbone of science. *Scientometrics, 64*(3), 351-374.

Bradford, S. C. (1934). Sources of information on specific subjects. *Engineering: An Illustrated Weekly Journal, 137*, 85-86.

Buter, R. K., Noyons, E. C. M., & Van Raan, A. F. J. (2010). Identification of converging research areas using publication and citation data. *Research Evaluation, 19*(1), 19-27.

Buter, R. K., Noyons, E. C. M., & Van Raan, A. F. J. (2011). Searching for converging research using field to field citations. *Scientometrics, 86*(2), 325-338.

Carvalho, R., & Batty, M. (2006). The geography of scientific productivity: Scaling in US computer science. *Journal of Statistical Mechanics: Theory and Experiment, 10*, DOI: 10.1088/1742-5468/2006/10/P10012

Cronin, B. & Pearson, S. (1990). The export of ideas from information science. *Journal of Information Science, 16*(6), 381-2391.

Cronin, B., & Davenport, L. (1989). Profiling the professors. *Journal of Information Science, 15*(1), 13-20.

Cronin, B., & Meho, L. I. (2008). The shifting balance of intellectual trade in information studies. *Journal of the American Society for Information Science & Technology, 59*(4), 551-564.

Garfield, E & Pudovkin, A. I, & Istomin, V. S. (2002). Algorithmic citation-linked historiography: Mapping the literature of science. *Proceedings of the American Society for Information Science and Technology Annual Meeting, 39*(1), 14-24.

Garfield, E. (1972). Citation analysis as a tool in journal evaluation. *Essays of an Information Scientist, 1*, 527-544. Retrieved February 27, 2012 from http://www.garfield.library.upenn.edu/essays/V1p527y1962-73.pdf?ref=Sawos.Org

Gazni, A., Sugimoto, C. R., & Didegah, F. (2011). Mapping world scientific collaboration: Authors, institutions, and countries. *Journal of the American Society for Information Science and Technology, 63*(2), 323-335.

Glänzel, W., & Moed, H. F. (2002). Journal impact measures in bibliometric research. *Scientometrics, 53*(2), 171-193.

Glänzel, W., & Schubert, A. (2003). A new classification scheme of science fields and subfields designed for scientometric evaluation purposes. *Scientometrics, 56*(3), 357-367.

Goldstone, R. L., & Leydesdorff, L. (2006). The import and export of cognitive science. *Cognitive Science, 30*(6), 983-993.

Guerrero-Bote, V. P., Zapico-Alonso, F. Z., Espinosa-Calvo, M. E., Gomez-Crisostomo, R., & Moya-Anegon, F. (2007). Import-export of knowledge between scientific subject categories: The iceberg hypothesis. *Scientometrics, 71*(3), 423-441.

Hessey, R. & Willett, P. (in press). Quantifying the value of knowledge exports from librarianship and information science research. *Journal of Information Science*. doi: 10.1177/0165551512442476

Jaffe, A. B., Trajtenberg, M., & Fogarty, M. S. (2000). Knowledge spillovers and patent citations: Evidence from a survey of inventors. *American Economic Review, 90*(2), 215-218.

Jaffe, A. B., Trajtenberg, M., & Henderson, A. D. (1993). Geographical localization of knowledge spillovers by patent citations. *Quarterly Journal of Economics, 108*(3), 577-599.

Janssens, F., Zhang, L., De Moor, B., & Glänzel, W. (2009). Hybrid clustering for validation and improvement of subject-classification schemes. *Information Processing & Management, 45*(6), 683-702.

Klavans, R., & Boyack, K. W. (2011). using global mapping to create more accurate document-level maps of research fields. *Journal of the American Society for Information Science and Technology, 62*(1), 1-18.

Klein, J. T. (1996). *Crossing boundaries: Knowledge, disciplinarities, and interdisciplinarities*. Charlottesville, VA: University Press of Virginia.

Larivière, V., Sugimoto, C. R., & Cronin, B. (2012). A bibliometric chronicling of Library and Information Science's first hundred years. *Journal of the American Society for Information Science and Technology*, *63*(5), 997-1016).

Lenoir, T. (1997). *Instituting science: The cultural production of scientific disciplines*. Stanford, CA: Stanford University Press.

Levitt, J. M., Thelwall, M., & Oppenheim, C. (2011). Variations between subjects in the extent to which the social sciences have become more interdisciplinary. *Journal of the American Society for Information Science and Technology, 62*(6), 1118-1129.

Lewison, G., Rippon, I., & Wooding, S. (2005). Tracking knowledge diffusion through citations. *Research Evaluation, 14*(1), 5-14.

Leydesdorff, L. (2011). "Structuration" by intellectual organization: the configuration of knowledge in relations among structural components in networks of science. *Scientometrics, 88*(2), 499-520.

Leydesdorff, L., & Probst, C. (2009). The delineation of an interdisciplinary specialty in terms of a journal set: The case of communication studies. *Journal of the American Society for Information Science and Technology, 60*(8), 1709-1718.

Leydesdorff, L., & Rafols, I. (2009). A Global map of science based on the ISI subject categories. *Journal of the American Society for Information Science and Technology, 60*(2), 348-362.

Leydesdorff, L., Carley, S., & Rafols, I. (in press). Global maps of science based on the new Web-of-Science categories. *Scientometrics*. doi: 10.1007/s11192-012-0783-8

Lillquist, E., & Green, S. (2010). The discipline dependence of citation statistics. *Scientometrics, 84*(3), 749-762.

Lockett, A., & McWilliams, A. (2005). The balance of trade between disciplines: do we effectively manage knowledge? *Journal of Management Inquiry, 14*(2), 139-150.

Lotka, A. J. (1926). The frequency distribution of scientific productivity. *Journal of the Washington Academy of Sciences, 16*, 317-324.

Mehta, A., Rysman, M., & Simcoe, T. (2010). Identifying the age profile of patent citations. *Social Science Research Network, 25*(7), 1179-1204.

Merton, R. K. (1968). The Matthew effect in science. *Science, 159*(3810), 56-63.

Morillo, F., Bordons, M., & Gomez, I. (2003). Interdisciplinarity in Science: A Tentative Typology of Disciplines and Research Areas. *Journal of the American Society for Information Science and Technology, 54*(13), 1237-1249.

Moya-Anegón, F., Vargas-Quesada, B., Herrero-Solana, V., Chinchilla-Rodríguez, Z., Corera-Álvarez, E., & Munoz-Fernández, F.J. (2004). A new technique for building maps of large scientific domains based on the cocitation of classes and categories. *Scientometrics, 61*(1), 129-145.

Nederhof, A. J. (2006). Bibliometric monitoring of research performance in the Social Sciences and the Humanities: A review. *Scientometrics, 66*(1), 81-100.

Nomaler, Ö., & Verspagen, B. (2008). Knowledge flows, patent citations and the impact of science on technology. *Economic Systems Research, 20*(4), 339-366.

Ponds, R., Van Oort, F., & Frenken, K. (2007). The geographical and institutional proximity of research collaboration. *Papers in Regional Science, 86*(3), 423-443.

Porter, A. L., & Rafols, I. (2009). Is science becoming more interdisciplinary? Measuring and mapping six research fields over time. *Scientometrics, 81*(3), 719-745.

Porter, A. L., Roessner, J. D., & Heberger, A. E. (2008). How interdisciplinary is a given body of research? *Research Evaluation, 17*(4), 273-282.

Porter, A. L., Roessner, J. D., Cohen, A. S., & Perreault, M. (2006). Interdisciplinary research: meaning, metrics and nurture. *Research Evaluation, 15*(3), 187-195.

Rafols, I., & Leydesdorff, L. (2009). Content-based and algorithmic classifications of journals: perspectives on the dynamics of scientific communication and indexer effects. *Journal of the American Society for Information Science and Technology, 60*(9), 1823-1835.

Rafols, I., & Meyer, M. (2010). Diversity and network coherence as indicators of interdisciplinarity: case studies in bionanoscience. *Scientometrics, 82*(2), 263-287.

Rafols, I., Porter, A. L., & Leydesdorff, L. (2010). Science overlay maps: A new tool for research policy and library management. *Journal of the American Society for Information Science and Technology, 61*(9), 1871-1887.

Redner, S. (1998). How popular is your paper? An empirical study of the citation distribution. *The European Physical Journal B - Condensed Matter and Complex Systems, 4*(2), 131-134.

Rinia E. J., Van Leeuwen, T. N., Bruins, E. E. W., Van Vuren, H. G., Van Raan, A. F. J. (2002). Measuring knowledge transfer between fields of science. *Scientometrics, 54*(3), 347-362.

Stigler, S. M. (1994). Citation patterns in the journals of statistics and probability. *Statistical Science, 9*(1), 94-108.

Suppes, P. (1984). *Probabilistic metaphysics*. Oxford, UK: Blackwell.

Van Leeuwen, T., & Tijssen, R. (2000). Interdisciplinary dynamics of modern science: analysis of cross-disciplinary citation flows. *Research Evaluation, 9*(3), 183-187.

Van Raan, A. F. J. (2008). Bibliometric statistical properties of the 100 largest European research universities: Prevalent scaling rules in the science system. *Journal of the American Society for Information Science and Technology, 59*(3), 461-475.

Wagner, C. S., & Leydesdorff, L. (2009). Network structure, self-organization and the growth of international collaboration in science. *Research Policy, 34*(10), 1608-1618.

Wouters, P. (1998). *The citation culture*. Ph.D. Thesis, University of Amsterdam, The Netherlands.

Xhignesse, L. V., & Osgood, C. E. (1967). Bibliographical citation characteristics of the psychological journal network in 1950 and in 1960. *American Psychologist, 22*(9), 778-791.

Yan, E., & Sugimoto, C. R. (2011). Institutional interactions: Exploring the social, cognitive, and geographic relationships between institutions as demonstrated through citation networks. *Journal of the American Society for Information Science and Technology, 62*(8), 1498-1514.

Zhang, L., Liu, X., Janssens, F., Liang, L., & Glänzel, W. (2010). Subject clustering analysis based on ISI category classification. *Journal of Informetrics, 4*(2), 185-193.

Zitt, M. (2005). Facing diversity of science: A challenge for bibliometric indicators. *Measurement, 3*(1), 38-49.